\documentclass[apl,twocolumn,notitlepage,reprint]{revtex4-1}

\usepackage{amsmath}
\usepackage{amssymb}
\usepackage{graphicx}
\usepackage{epstopdf}
\usepackage[colorlinks=true]{hyperref}
\DeclareMathOperator{\sgn}{sgn}

\begin{document}
\title{Fast Vortex Oscillations in a Ferrimagnetic Disk near the Angular Momentum Compensation Point}

\author{Se Kwon Kim}
\affiliation{Department of Physics and Astronomy, University of California, Los Angeles, California 90095, USA}

\author{Yaroslav Tserkovnyak}
\affiliation{Department of Physics and Astronomy, University of California, Los Angeles, California 90095, USA}

\date{\today}

\begin{abstract}
We theoretically study the oscillatory dynamics of a vortex core in a ferrimagnetic disk near its angular momentum compensation point, where the spin density vanishes but the magnetization is finite. Due to the finite magnetostatic energy, a ferrimagnetic disk of suitable geometry can support a vortex as a ground state similar to a ferromagnetic disk. In the vicinity of the angular momentum compensation point, the dynamics of the vortex resemble those of an antiferromagnetic vortex, which is described by equations of motion analogous to Newton's second law for the motion of particles. Owing to the antiferromagnetic nature of the dynamics, the vortex oscillation frequency can be an order of magnitude larger than the frequency of a ferromagnetic vortex, amounting to tens of GHz in common transition-metal based alloys. We show that the frequency can be controlled either by applying an external field or by changing the temperature. In particular, the latter property allows us to detect the angular momentum compensation temperature, at which the lowest eigenfrequency attains its maximum, by performing FMR measurements on the vortex disk. Our work proposes a ferrimagnetic vortex disk as a tunable source of fast magnetic oscillations and a useful platform to study the properties of ferrimagnets.
\end{abstract}

\maketitle

\begin{figure}
\includegraphics[width=0.9\columnwidth]{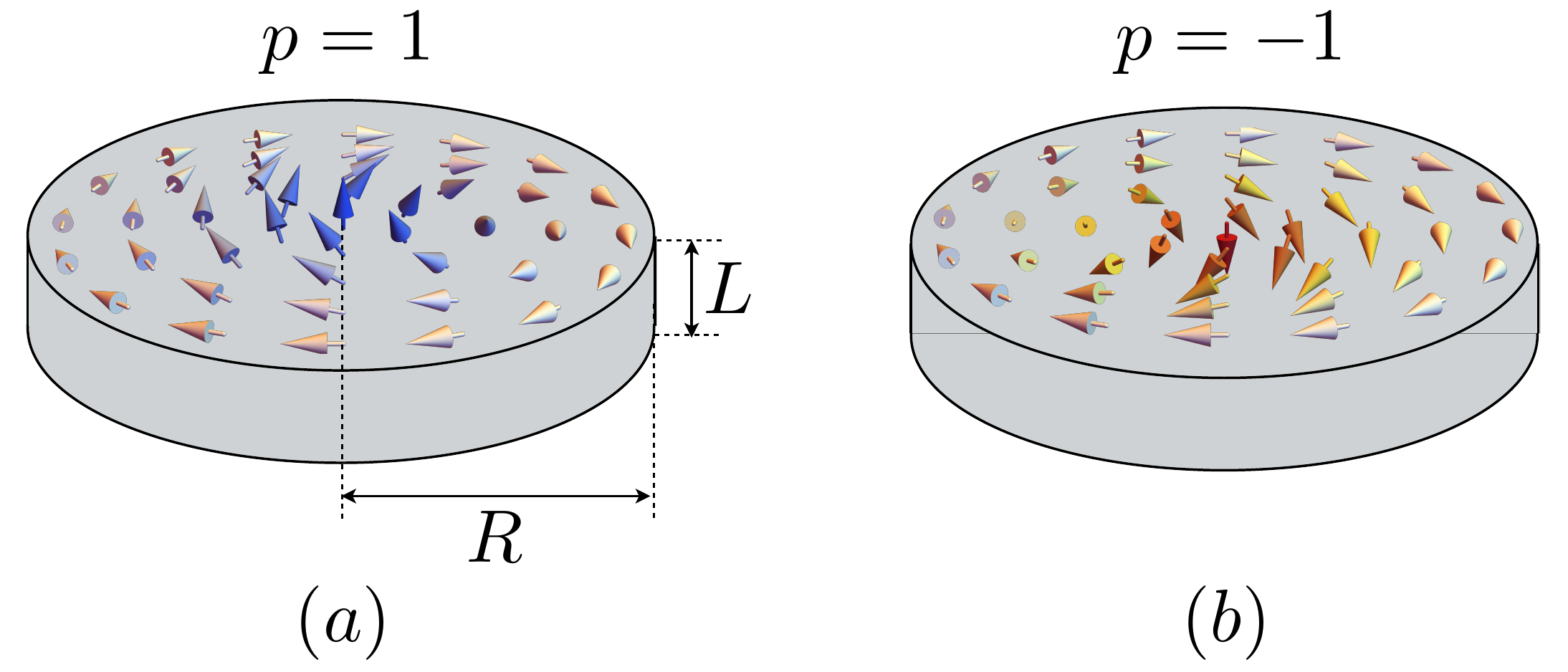}
\caption{Schematic illustrations of the magnetization of ferrimagnetic disks harboring (a) a vortex with polarization $p = 1$ and (b) a vortex with polarization $p = -1$. The thickness and the radius of disks are denoted by $L$ and $R$, respectively.}
\label{fig:fig1}
\end{figure}

Ferromagnets are used as platforms to produce GHz magnetic oscillations, e.g., spin-torque and spin-Hall oscillators~\cite{ChenPIEEE2016}, which can be used as microwave signal generators. Among them, the core oscillation of a vortex in a microdisk~\cite{CowburnPRL1999, *NovosadPRB2002, *MetlovJMMM2002, *IvanovAPL2002, ParkPRB2003}, which is a curling magnetization texture stabilized by magnetostatic energy, stands out because of its high coherence~\cite{KasaiPRL2006, PribiagNP2007, *DussauxNC2010}. This advantage allows it to be used in areas requiring highly precise oscillations such as biomedicine~\cite{KimNM2010, *VitolN2012, *Goiriena-GoikoetxeaN2016}. Aside from the dynamics of an isolated vortex, coupled oscillations of vortices in an array of disks have been studied as magnonic vortex crystals~\cite{ShibataPRB2003, *ShibataPRB2004, HanSR2013, *BehnckePRB2015, *StreubelPRB2015, *HanzeSR2016}. The oscillation can be driven in several ways such as with an external field~\cite{GuslienkoJAP2002, YuSR2013} or with an electric current~\cite{KasaiPRL2006, PribiagNP2007}. Its characteristic frequency is given by 
\begin{equation}
\label{eq:wfm}
f_\text{FM} \sim \gamma \mu_0 M_s \, ,
\end{equation}
up to the geometric factor, where $\gamma$ is the gyromagnetic ratio, $\mu_0$ is the permeability constant, and $M_s$ is the saturation magnetization. The observed frequency ranges from several hundred MHz up to $2$ GHz~\cite{ParkPRB2003, VogelPRL2010, *SugimotoPRL2011, PribiagNP2007, YuSR2013}. See Fig.~\ref{fig:fig1} for some schematic illustrations of vortex disks.

Recently, antiferromagnets have been gaining attention as possible hosts of magnetic oscillations that are orders-of-magnitude faster than their ferromagnetic counterparts~\cite{*[][{, and references therein.}] GomonayLTP2014, *[][{, and references therein.}] JungwirthNN2016}. For example, a spin Hall oscillator based on an antiferromagnet/heavy-metal heterostructure has been proposed as a possible THz signal generator~\cite{ChengPRL2016, *KhymynSR2017, *ZarzuelaPRB2017}. However, the possibility of fast oscillations of an antiferromagnetic vortex has not been investigated because, differing from ferromagnetic cases, antiferromagnetic disks do not support a vortex as an equilibrium state due to the absence of magnetostatic energy.

Here, we show that a ferrimagnetic disk can host a stable vortex whose oscillations can be orders-of-magnitude faster than those of ferromagnetic vortices. A ferrimagnetic disk can harbor a vortex as a ground state due to the finite magnetostatic energy~\cite{*[][{, and references therein.}] KirilyukRPP2013}, which confines the vortex core to the center of the disk. We show that at the angular momentum compensation point (CP), where the spin density vanishes, the nature of the vortex-core dynamics is antiferromagnetic and in this way the core behaves like a classical particle in a potential well~\cite{IvanovPRL1994, *DasguptaPRB2017}. The characteristic frequency of an oscillation of the vortex core at the CP, which will be derived below, is given by 
\begin{equation}
\label{eq:wafm}
f_\text{AFM} \sim \sqrt{ \frac{J}{\hbar} * \gamma_t \mu_0 M_s} \, ,
\end{equation}
up to a dimensionless geometric factor. Here, $J$ is the antiferromagnetic exchange energy between neighboring magnetic moments and $\gamma_t$ is the transverse gyromagnetic ratio with respect to the direction of the magnetization. We will show that this frequency is in the range of tens of GHz, e.g., $f_\text{AFM} \approx 30$ GHz for the material parameters of CoTb~\cite{FinleyPRA2016}. As a fast-oscillating source, a ferrimagnetic vortex in a microdisk has two advantages over other proposals using antiferromagnets. First, it can be easily operated by an external magnetic field due to the finite magnetization. Secondly, since the spin density can be controlled by changing the temperature~\cite{StanciuPRB2006}, the oscillation frequency can be thermally tuned. In particular, the latter advantage allows us to detect the angular momentum compensation temperature, where the lowest frequency will exhibit its peak, by performing FMR measurements. The high coherence of the vortex oscillation~\cite{KasaiPRL2006, PribiagNP2007} will yield a narrow FMR linewidth. A recent realization of a vortex in an artificial-ferrimagnet microdisk composed of Py/Gd superlattices~\cite{LapaarXiv2017} supports the experimental feasibility of our proposal.


Let us begin by deriving the equations of motion for a vortex in a ferrimagnetic disk. We will describe the low-energy dynamics of a ferrimagnet using coarse-grained variables over the constituent sublattices, following the approach taken by \textcite{AndreevSPU1980} and subsequently used in Refs.~\cite{IvanovSSC1984, *LossPRL1992, *BelashchenkoAPL2016, KimPRB2017, ChioleroPRB1997}. This approach is based on approximate exchange symmetry and does not use any model representations of the state of the magnet, e.g., the number of sublattices as pointed in Ref.~\cite{AndreevSPU1980}. The collinear order of a ferrimagnet much below the Curie temperature can be described by the unit vector $\mathbf{n}$ denoting its direction. Along this direction, a ferrimagnet has the net magnetization, $\mathbf{M} = M_s \mathbf{n}$, and the net spin density, $\mathbf{s} = s \mathbf{n}$, which are related by the (longitudinal) gyrotropic coefficient $\gamma$ as $\mathbf{M} = \gamma \mathbf{s}$. Here, $M_s$ is the net saturation magnetization density. There is a class of ferrimagnets such as rare-earth transition-metal alloys, which exhibit two distinct special values of temperature or chemical composition: the angular momentum CP, where the spin density vanishes, $s = 0$, and the magnetization CP, where the magnetization vanishes, $M_s = 0$~\cite{KirilyukRPP2013}. These two points are different due to the distinct gyromagnetic ratios of constituent sublattices. In particular, at the angular momentum CP, these ferrimagnets have the finite magnetization and the zero spin density~\cite{StanciuPRB2006}, allowing the unconventional coexistence of magnetostatic energy and antiferromagnetic dynamics. The former stabilizes a vortex in the microdisk and the latter yields fast oscillations of the vortex, as will be explained below.

The dynamics of a generic collinear ferrimagnet can be described by the following Lagrangian density~\cite{IvanovSSC1984, KimPRB2017}:
\begin{equation}
\label{eq:L}
\mathcal{L} = - s \mathbf{a} [\mathbf{n}] \cdot \dot{\mathbf{n}} + \rho \dot{\mathbf{n}}^2 / 2 - \mathcal{U}[\mathbf{n}] \, ,
\end{equation}
where $\mathbf{n}$ is the unit vector in the direction of the local magnetization, $\mathbf{a}$ is the vector potential for the magnetic monopole, $\boldsymbol{\nabla}_{\mathbf{n}} \times \mathbf{a} = \mathbf{n}$, and $\rho$ is the effective inertia density. Here, the first term represents the Berry phase term associated with the spin density; the second term represents the inertia of the dynamics of $\mathbf{n}$ which can arise due to the relative canting of sublattice spins; the third term is the free-energy density. This Lagrangian for ferrimagnets can be interpreted as a hybrid of those for ferromagnets and antiferromagnets. We model the free-energy density in the absence of an external field by
\begin{equation}
\label{eq:U}
\mathcal{U} = \frac{A (\boldsymbol{\nabla} \mathbf{n})^2 + \mu_0 \mathbf{H}_\text{ms}^2}{2} \, .
\end{equation}
Here, the first term is the exchange energy parametrized by the stiffness coefficient $A > 0$; the second term is the magnetostatic energy, where $\mathbf{H}_\text{ms}$ is the dipolar field induced by the magnetization satisfying the equations $\boldsymbol{\nabla} \cdot (\mathbf{H}_\text{ms} + \mathbf{M}) = 0$ and $\boldsymbol{\nabla} \times \mathbf{H}_\text{ms} = 0$. At the angular momentum CP, $s = 0$, the kinetic part of the Lagrangian is antiferromagnetic, but the magnetostatic energy is finite owing to the finite magnetization, $M_s \neq 0$. Recognizing this peculiar coexistence of antiferromagnetic dynamics and magnetostatic energy motivated this work. The energy dissipation associated with the dynamics can be accounted for by considering the Rayleigh dissipation function, $\mathcal{R} = s_\alpha \dot{\mathbf{n}}^2 / 2$, which is half of the energy (density) dissipation rate. The parameter $s_\alpha$ is reduced to the product of the Gilbert damping constant~\cite{GilbertIEEE2004} and the spin density in the ferromagnetic limit, $\rho \rightarrow 0$.

There is a length scale associated with the energy density $\mathcal{U}$, which is given by $\lambda \equiv \sqrt{A / \mu_0 M_s^2}$. It is referred to as the exchange length at which the magnetostatic energy is comparable to the exchange energy. When a dimension of the sample is larger than the exchange length, the magnetostatic energy dominates the total energy and generates nonuniform ground states. Let us consider a circular ferrimagnetic disk of radius $R$ and thickness $L$ as shown in Fig.~\ref{fig:fig1}. When the thickness is order of the exchange length, $L \sim \lambda$, and the aspect ratio of the thickness to the radius is small enough, $g \equiv L / R \lesssim 0.5$, the disk supports a vortex as a ground state~\footnote{The phase diagram can be found in Ref.~\cite{HaPRB2003}.}. For a quantitative estimate of the geometry of the physical system, let us take the example of Co$_{1-x}$Tb$_x$, which exhibits its angular momentum CP at $x \approx 0.17$. According to the experimental results in Ref.~\cite{FinleyPRA2016}, the parameters are given by $A \approx 1.4 \times 10^{-11}$ J/m and $M_s \approx 10^5$ A/m at the angular momentum CP, which yields the exchange length $\lambda \approx 30$ nm. Therefore, for example, a disk made of CoTb with thickness $L = 100$ nm and radius $R = 1000$ nm would have a vortex as a ground state.

The observed vortex states of magnetic disks can be described by the following ansatz in the spherical-coordinate representation, $\mathbf{n} = (\sin \theta \cos \phi, \sin \theta \sin \phi, \cos \theta)$: $\theta = (1 - p) \pi / 2 + 2 \arctan(\sqrt{x^2 + y^2} / R_c)$ for $\sqrt{x^2 + y^2} < R_c$ and $\theta = \pi / 2$ otherwise, and $\phi = \varphi + c \pi / 2$, where $R_c$ is the vortex-core radius that is on the order of the exchange length $\lambda$~\cite{GuslienkoAPL2001}. Here, $p = \pm 1$ is the direction of the magnetization at the core, $\mathbf{n} = p \hat{\mathbf{z}}$, which is referred to as the polarization; $c = \pm 1$ corresponds to the counter-clockwise ($+$) or clockwise ($-$) rotation of the magnetization in the plane, which is referred to as the chirality.  
The low-energy dynamics of a vortex can be described by the dynamics of its core position, $\mathbf{R}(t) = (X, Y)$, with the aforementioned ansatz. The equations of motion for the position can be derived from the above Lagrangian and the Rayleigh dissipation function within the collective-coordinate approach, assuming a rigid magnetization profile~\cite{ThielePRL1973, *TretiakovPRL2008, KimPRB2017}:
\begin{equation}
\label{eq:eom}
M \ddot{\mathbf{R}} + G \dot{\mathbf{R}} \times \hat{\mathbf{z}} + D \dot{\mathbf{R}} = - K \mathbf{R} \, ,
\end{equation}
where 
\begin{subequations}
\label{eq:coeff}
\begin{align}
M &= 2 \pi \rho L \ln (R / R_c) \, \\
G &= 2 \pi p s L \, ,\label{eq:G} \\ 
D &= 2 \pi s_\alpha L \ln(R / R_c) \, , \\
K &= \mu_0 L M_s^2 [F_1(L/R) - (\lambda / R)^2] \, .
\end{align}
\end{subequations}
Here, $M$ represents the mass of the vortex, which originates from the inertial term in the Lagrangian; $G$ parametrize the gyrotropic force on the vortex, which is rooted in the spin Berry phase; $D$ parametrizes the viscous force on the vortex. The right-hand side is the restoring force on the vortex, which has been obtained in Ref.~\cite{GuslienkoJAP2002} for ferromagnetic vortices. When the vortex core is away from the center of the disk, magnetostatic charges are created on the boundary. The corresponding magnetostatic energy engenders a confining potential for the vortex core, which is parametrized by $K$.  The factor $[F_1(L/R) - (\lambda / R)^2]$ is a dimensionless number that is on the order of 0.1 for $L/R \sim 0.1$. The definition of the function $F_1(x)$, which is an increasing function of $x$, can be found in Ref.~\footnote{The function $F_1$ is defined as follows: $F_1(g) = \int_0^\infty dk f(k g) J_1^2 (k) / k$ with $f(x) = 1 - (1 - \exp(-x))/x$, where $J_1 (x)$ is the Bessel function of the first kind~\cite{GuslienkoJAP2002, ShibataPRB2003}.}. Note that the gyrotropic coefficient is proportional to the spin density, $G \propto s$, and thus it vanishes at the angular momentum CP.

\begin{figure}
\includegraphics[width=0.8\columnwidth]{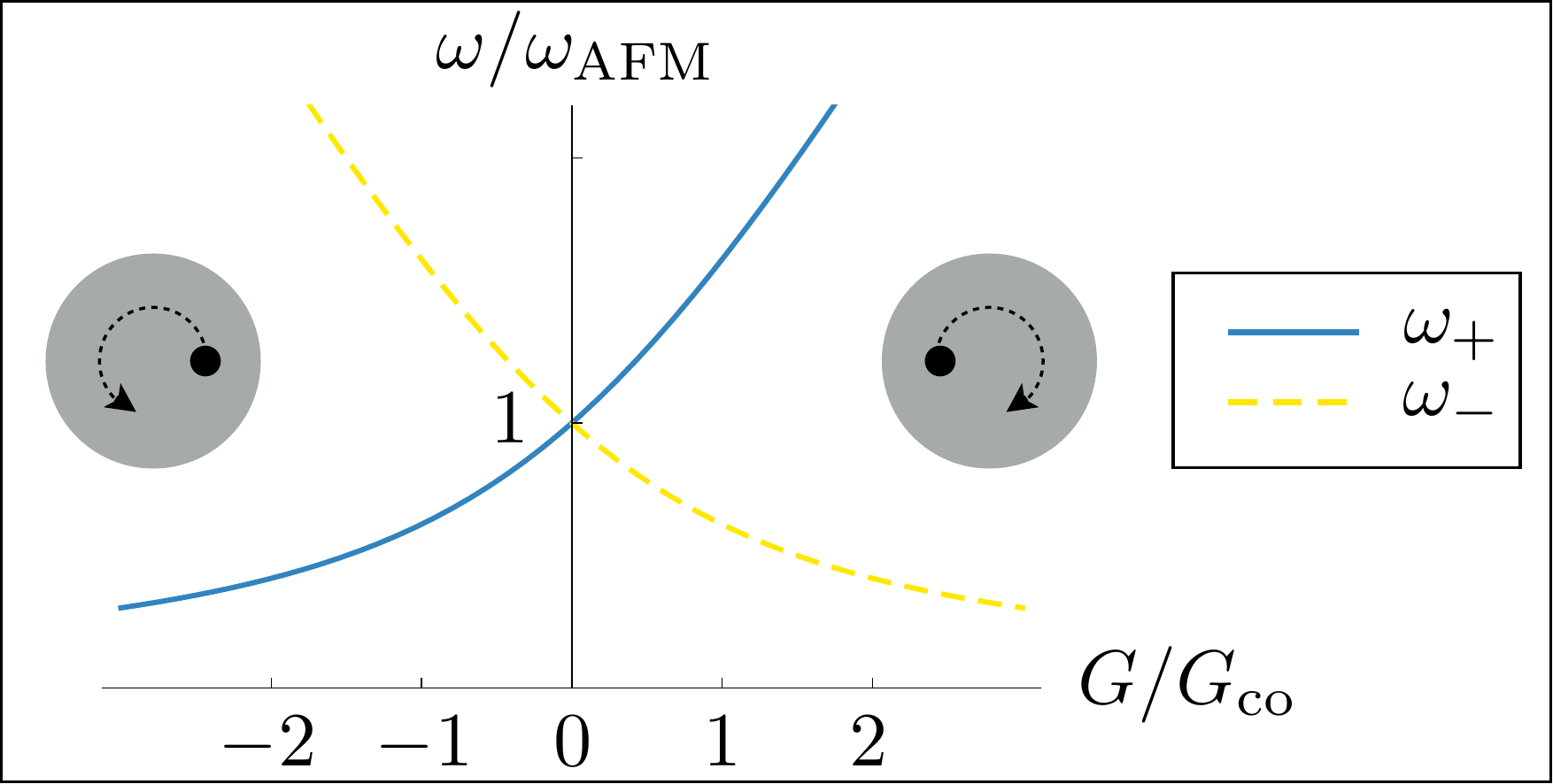}
\caption{The eigenfrequencies $\omega_\pm / \omega_\text{AFM}$ as functions of the gyrotropic coefficient $G / G_\text{co}$. The solid and dashed lines show the eigenfrequencies for the counter-clockwise $(\omega_+)$ and clockwise $(\omega_-)$ rotations of the core. See the main text for discussions.}
\label{fig:fig2}
\end{figure}

Let us study the excitation mode for the core dynamics described by the above equation. To this end, it is convenient to express the equations of motion in terms of the complex variable, $\Psi \equiv X + i Y$:
\begin{equation}
\label{eq:eom2}
M \ddot{\Psi} - i G \dot{\Psi} + D \dot{\Psi} = - K \Psi \, .
\end{equation}
When the damping constant is small, $\alpha \ll 1$, the main effect of the viscous force is to broaden the linewidth of the excitation spectrum, not to change the spectrum itself, and so we will neglect it henceforth by setting $D = 0$. The oscillation frequencies of the monochromatic solution, $\Psi(t) \propto \exp(i \omega t)$, are given by
\begin{equation}
\label{eq:sol}
\omega_\pm = \frac{G}{2M} \pm \sqrt{ \left( \frac{G}{2 M} \right)^2 + \frac{K}{M} }\, .
\end{equation}
At the angular momentum CP, where the gyrotropic force vanishes $G = 0$, the eigenfrequencies are given by $\omega_\pm = \pm \omega_\text{AFM}$ with $\omega_\text{AFM} \equiv \sqrt{K / M}$, which is reduced to Eq.~(\ref{eq:wafm}) if we recast it in terms of the microscopic parameters using $\rho \sim \hbar^2 / J a^3$ with $a$ the lattice constant. The absence of any gyrotropic coupling between $X$ and $Y$ is one characteristic of the antiferromagnetic dynamics of two-dimensional solitons~\cite{TvetenPRL2013} such as vortices~\cite{IvanovPRL1994} and skyrmions~\cite{BarkerPRL2016}. Note that two circularly polarized modes are degenerate at the CP, where the spin configurations respect time-reversal symmetry. Far away from the angular momentum CP, where the gyrotropic force dominates the dynamic part in the equations of motion, the lowest eigenfrequencies are given by $\omega = - p \sgn(\gamma) \omega_\text{FM}$ with $\omega_\text{FM} \equiv K / |G|$, which corresponds to the ferromagnetic case~\cite{GuslienkoJAP2002}. The crossover between antiferromagnetic and ferromagnetic dynamics occurs when the two frequencies are comparable, $\omega_\text{FM} \sim \omega_\text{AFM}$, corresponding to $|G| \sim G_\text{co} \equiv \sqrt{M K}$. The above equation (\ref{eq:sol}) for general cases can be written as a function of the gyrotropic coefficient $G$: $\omega_\pm / \omega_\text{AFM} = G / 2 G_\text{co} \pm \sqrt{(G / 2 G_\text{co})^2 + 1}$, which is shown in Fig.~\ref{fig:fig2}.

Let us provide a numerical estimate for the antiferromagnetic oscillation at the angular momentum CP. The alloy Co$_{1-x}$Tb$_x$ at its CP, $x \approx 17$, has the exchange-stiffness coefficient, $A \approx 1.4 \times 10^{-11}$ J/m, and the lattice constant, $a \approx 0.4$ nm, which yield the microscopic exchange constant $J = A a \approx 35$ meV~\cite{KirilyukRPP2013}. By using the inertia, $\rho = \hbar^2 / 2 J z a^3$~\cite{ChioleroPRB1997, KimPRB2014}, where $z = 6$ is the coordination number for three-dimensional bipartite lattices, and the additional parameters, $L/R = 0.1$ and $R_c \approx \lambda$, we can estimate the eigenfrequency at the CP: $f_\text{AFM} \equiv \omega_\text{AFM} / 2 \pi \approx 30$ GHz, which is one order-of-magnitude larger than the observed frequencies in ferromagnetic disks of several hundred MHz up to $2$ GHz~\cite{ParkPRB2003, VogelPRL2010, PribiagNP2007, YuSR2013}. For example, for a cobalt disk of the same shape, the ferromagnetic resonance frequency is calculated as $f_\text{FM} \equiv \omega_\text{FM} / 2 \pi \approx 700$ MHz when using the saturation magnetization $M_s \approx 1.2 \times 10^6$ A/m measured for $30$-nm-thick films~\cite{WangAPL2009, *MruczkiewiczPRB2012}.

\begin{figure}
\includegraphics[width=0.9\columnwidth]{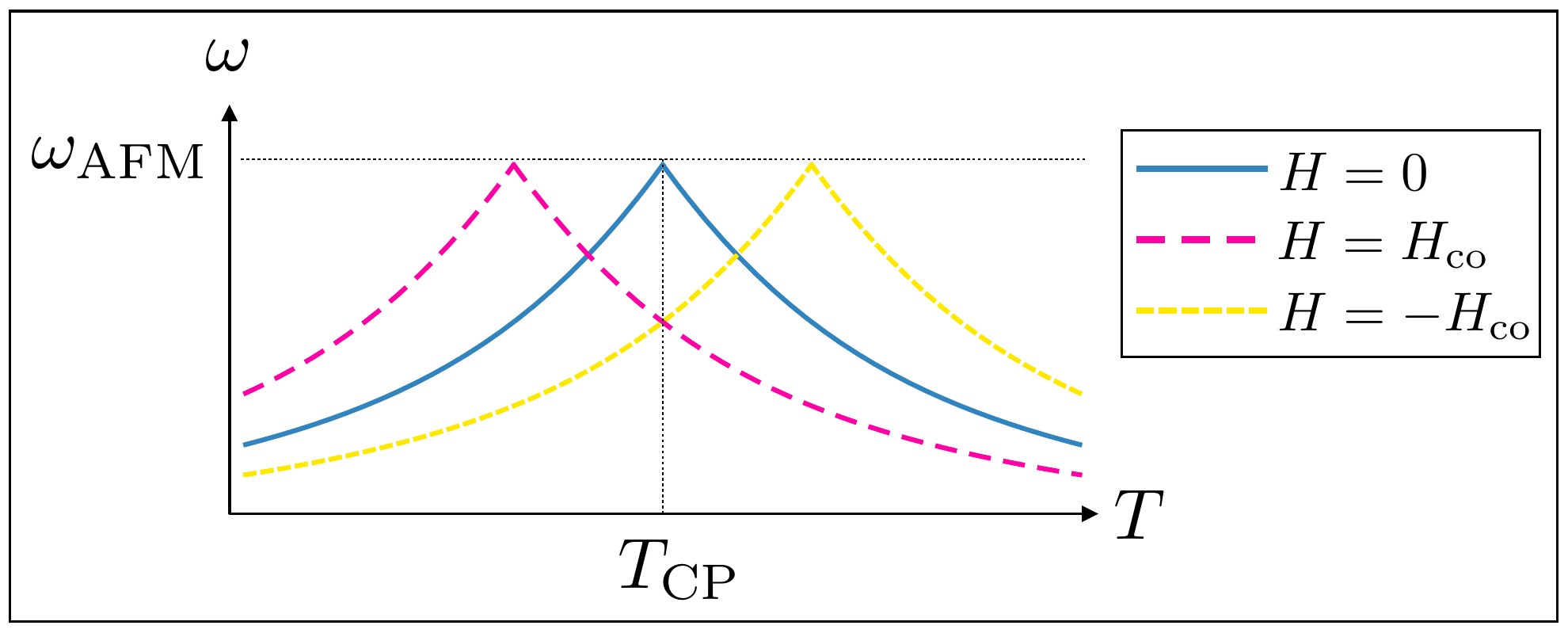}
\caption{Schematic illustrations of the lowest eigenfrequency $\omega$ of the vortex-core oscillation as a function of temperature $T$, in the vicinity of the angular momentum CP ($T_\text{CP}$) subjected to an external field, $\mathbf{H} = H \hat{\mathbf{z}}$. The vortex with polarization $p = 1$ is considered. For CoTb disks, the maximum eigenfrequency and the crossover field are estimated as $\omega_\text{AFM} \approx 2 \pi \times 30$ GHz and $H_\text{co} \approx 2$ T, respectively. See the main text for discussions.}
\label{fig:fig3}
\end{figure}

The dependence of the eigenfrequency on the gyrotropic coefficient can be used to infer the angular momentum CP. For example, when we measure the ferromagnetic resonance (FMR) frequency of the vortex oscillation by varying the temperature across the angular momentum CP denoted by $T_\text{CP}$, the lowest resonance frequency should attain its maximum at $T_\text{CP}$. In addition, since the rotational directions of the core oscillation below and above $T_\text{CP}$ are opposite, $T_\text{CP}$ can be measured by detecting the change of the oscillation direction, which can be probed by time-resolved scanning transmission X-ray microscopy~\cite{HanSR2013}. See Fig.~\ref{fig:fig3} for schematic illustrations of the lowest oscillation frequency as a function of a temperature. These methods using a vortex oscillation in a ferrimagnetic disk to determine $T_\text{CP}$ can be an alternative to a recent proposal based on domain-wall speed measurements~\cite{KimarXiv2017}.


We now study the effects of an external field applied perpendicular to the disk. In Refs.~\cite{IvanovPRL1994}, it has been shown that the application of an external field can induce a gyrotropic force on antiferromagnetic solitons. The same mechanism works for ferrimagnets. In the presence of an external field, $\mathbf{H}$, the inertial term in the Lagrangian~(\ref{eq:L}) is changed to $\rho (\dot{\mathbf{n}} - \gamma_t \mathbf{n} \times \mathbf{H})^2 / 2$~\cite{AndreevSPU1980}. The field-induced dynamical term, $- \rho \gamma_t \dot{\mathbf{n}} \cdot (\mathbf{n} \times \mathbf{H})$, is linear in the time derivative of $\mathbf{n}$, and thus can be interpreted as the effective geometric phase associated with the magnetic dynamics. When the field is perpendicular to the disk, $\mathbf{H} = H \hat{\mathbf{z}}$, it gives rise to an extra force term in the equations of motion~(\ref{eq:eom}) for a vortex within the collective coordinate approach~\cite{IvanovPRL1994}:
\begin{equation}
M \ddot{\mathbf{R}} + G \dot{\mathbf{R}} \times \hat{\mathbf{z}} + G_H \dot{\mathbf{R}} \times \hat{\mathbf{z}} + D \dot{\mathbf{R}} = - K \mathbf{R} \, ,
\end{equation}
where
\begin{equation}
G_H \equiv 2 \pi p \rho \gamma_t L H
\end{equation}
is the contribution to the gyrotropic coefficient induced by an external field. Its origin can be understood by writing $\rho \gamma_t H$ in the field-induced kinetic term as $\chi H / \gamma_t$ by using the relation, $\rho = \chi / \gamma_t^2$~\cite{ChioleroPRB1997}, where $\chi$ is the transverse magnetic susceptibility. This induced spin density contributes to the gyrotropic coefficient $G$ [Eq.~(\ref{eq:G})]. The crossover field corresponding to the crossover gyrotropic coefficient is given by $H_\text{co} = G_\text{co} / 2 \pi \rho \gamma_t L$, which is estimated to be $H_\text{co} \approx 2$T for the aforementioned CoTb disk. The field-induced gyrotropic force can be inferred by performing FMR measurements. See Fig.~\ref{fig:fig3} for schematic illustrations~\footnote{According to Ref.~\cite{KirilyukRPP2013}, the angular momentum CP is typically $50$ K above the magnetization CP in rare-earth transition-metal ferromagnetic alloys. It implies that the (longitudinal) gyromagnetic ratio is negative below the magnetization CP and above the angular momentum CP, and positive in between. In Fig.~\ref{fig:fig3}, we used this dependence of the gyromagnetic-ratio sign on the temperature around the angular momentum CP in order to determine how the excitation modes are displaced for finite fields.}.


To conclude, we have studied the oscillation dynamics of a vortex core in a ferrimagnetic disk. We have shown that the oscillation frequency can be orders-of-magnitude larger than that of a ferromagnetic vortex in the vicinity of the angular momentum CP, where the nature of the dynamics is antiferromagnetic due to the vanishing spin density. This is similar to the recently observed enhancement of the ferrimagnetic domain-wall speed around the angular momentum CP~\cite{KimarXiv2017}. The vortex oscillation frequency can be tuned by changing the temperature or by applying an external field. Our results exemplify the possibility of using ferrimagnets to realize desired functionalities that have been difficult to achieve with conventional ferro- and antiferromagnets. One possible research topic related to this is the dynamics of a vortex domain wall in a long ferrimagnetic strip, which can be faster than its ferromagnetic counterpart in the vicinity of the angular momentum CP and thus may realize a better racetrack memory~\cite{ParkinScience2008}. 

We made several approximations in this work to describe the dynamics of a ferrimagnetic vortex. First, we obtained the mass of a vortex by assuming that its magnetization profile is rigid. There can be contributions to the mass from deformations of the profile as shown for ferromagnetic magnetic bubbles~\cite{MakhfudzPRL2012}. Secondly, we neglected the magnetic crystalline anisotropy by assuming that the magnetostatic energy dominates it. We remark that rare-earth transition-metal ferromagnetic alloys are known to have uniaxial crystalline anisotropy. The anisotropy type can be easy axis or easy plane, depending on, e.g., the deposition parameters controlling the structure of the film~\cite{HansenJAP1989, *OstlerNC2010}. Thirdly, the derived equations of motion for the dynamics of a vortex are valid to linear order in the vortex-core displacement. There can arise nonlinear effects beyond our description when the amplitude of oscillations is sufficiently large. Fourthly, when studying the temperature dependence of the eigenfrequencies across the angular momentum CP, we have assumed that the material parameters other than the net spin density, such as the effective inertia or the exchange stiffness, are constant. Smooth variations of those parameters should not affect the dynamics of a vortex significantly.


We thank Oleg Tchernyshyov and Ricardo Zarzuela for discussions on ferromagnetic vortices and comments on the manuscript, and Kyung-Jin Lee for discussions on physical realizations of ferrimagnetic vortex disks. This work was supported by the Army Research Office under Contract No. W911NF-14-1-0016.

%

\end{document}